\documentclass[
    ,final            
  ]
  {aipproc}

\layoutstyle{8x11double}

\begin{document}

\title{Prolonged activity of the central engine of Gamma Ray Bursts}


\author{G. Ghisellini}{
  address={INAF--Osservatorio Astronomico di Brera, Via E. Bianchi, 46 I--23807, Merate, Italy}
}

\author{M. Nardini}{
	address={SISSA/ISAS, Via Beirut 2--4, I--34151, Trieste, Italy},
}

\author{G. Ghirlanda}{
  address={INAF--Osservatorio Astronomico di Brera, Via E. Bianchi, 46 I--23807, Merate, Italy}
}

\begin{abstract}

We call ``prompt" emission of Gamma Ray Bursts (GRBs) the erratic and
violent phase of hard X-ray and soft $\gamma$--ray emission, 
usually lasting for tens of seconds in long GRBs.
However, the central engine of GRBs may live much longer.
Evidence of it comes from the strange behaviour
of the early ``afterglow", seen especially in the X--ray band,
characterised by a ``steep--flat--steep" light--curve,
very often not paralleled by a similar behaviour in the
optical band.
This difference makes it hard to explain {\it both} the optical
and the X--ray emission with a unique component.
Two different mechanisms seem to be required.
One can well be the standard emission from the forward shock 
of the fireball running through the interstellar medium.
The second one is more elusive, and to characterise its properties 
we have studied those GRBs with well sampled data and known redshift, 
fitting {\it at the same time} the optical and the X--ray light--curves 
by means of a composite model, i.e. the sum of the standard 
forward shock emission and a phenomenological additional component 
with a minimum of free parameters.
Some interesting findings have emerged, pointing to a long
activity of the central engine powered by fallback material.
 
\end{abstract}

\maketitle


\section{Introduction}

The {\it Swift} satellite \cite{gehrels04}
certainly revolutionised the GRB field,
but it has also shaken some of our pre--{\it Swift} strong beliefs.
One of those was the reassuring division between the prompt 
and the afterglow phases understood in terms of internal shocks
between relativistic shells followed by the external shock of the fireball
with the circumburst medium.
This scenario could well explain the erratic and violent prompt
emission lasting up to a few tens of seconds, and most of the
(then existing) observations of the afterglow, typically
starting a few hours after the trigger.
The afterglow phase was thought to be somewhat better understood,
since it was constructed upon the very solid basis
of the conservation of energy and momentum, shock acceleration
physics and the synchrotron process as the main radiation mechanism
(that was even confirmed by the detection of a small, but significant,
linear polarisation; see \cite{covino09} for a recent review).
Although some problem remained to be solved for the prompt phase
(as the inherently low efficiency of internal shocks 
and the hardness of the
observed spectra if the emission process is synchrotron by
cooling electrons, \cite{gg00}), the overall scenario was satisfying,
especially after the discovery that long GRBs were indeed
associated with powerful supernovae, confirming that the
central engine had to be an hyper--accreting stellar size black hole.

The {\it Swift} capability of a very fast slew allowed to explore 
the very early phases, soon discovering the unexpected 
``Steep--Flat--Steep" behaviour of the light--curve, 
especially in the X-rays \cite{gt05, nousek05}.
Although interpreted in several ways
(see e.g. \cite{zhang07a} for a recent review),
none  seems conclusive.
The spectral slope does not change across the temporal
break from the shallow to the normal decay phase,
ruling out a changing spectral break as a viable explanation.
An hydrodynamical or geometrical nature of the break is instead preferred.

\begin{figure}
  \includegraphics[height=8.5cm]{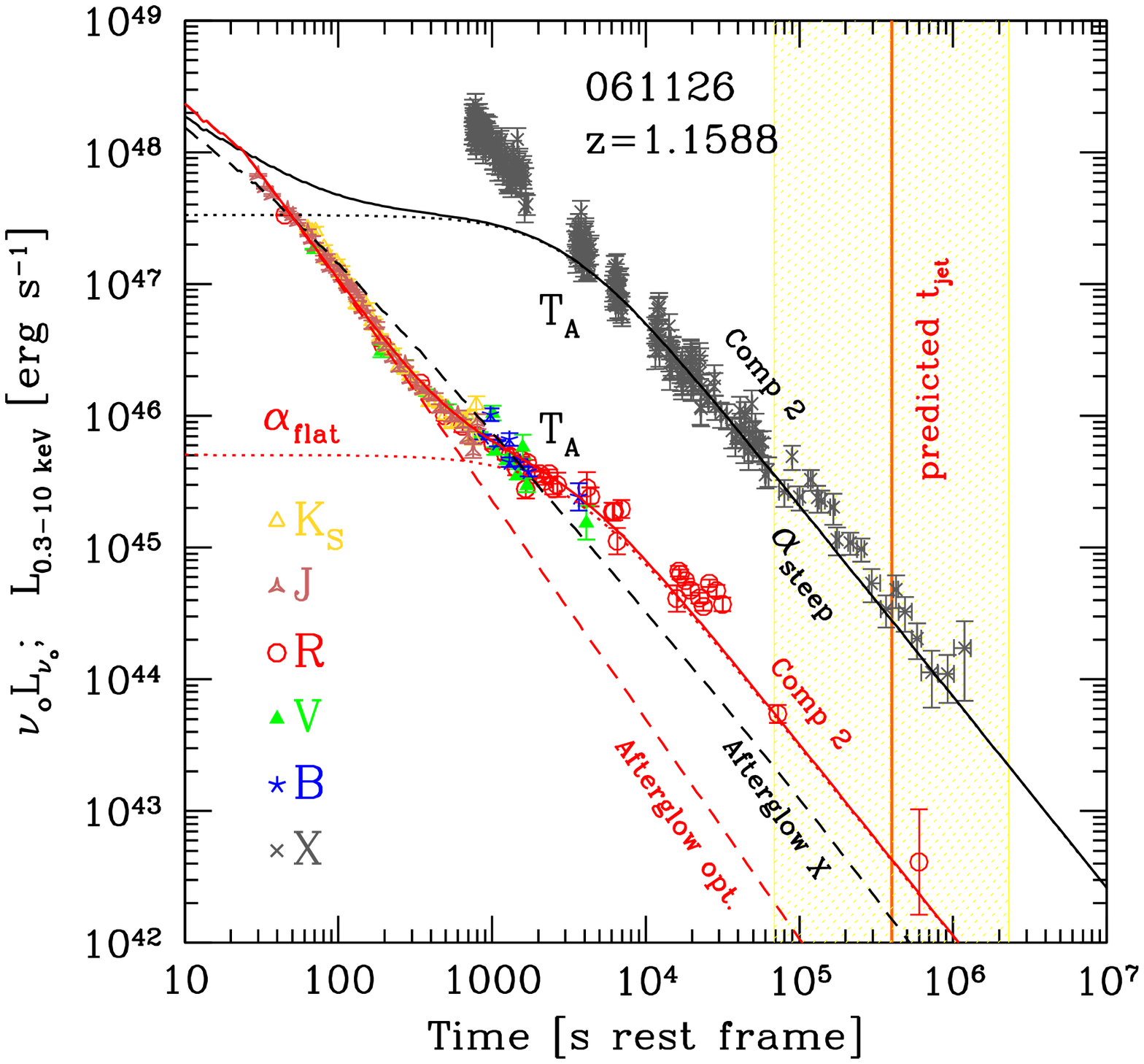} 
  \includegraphics[height=8.5cm]{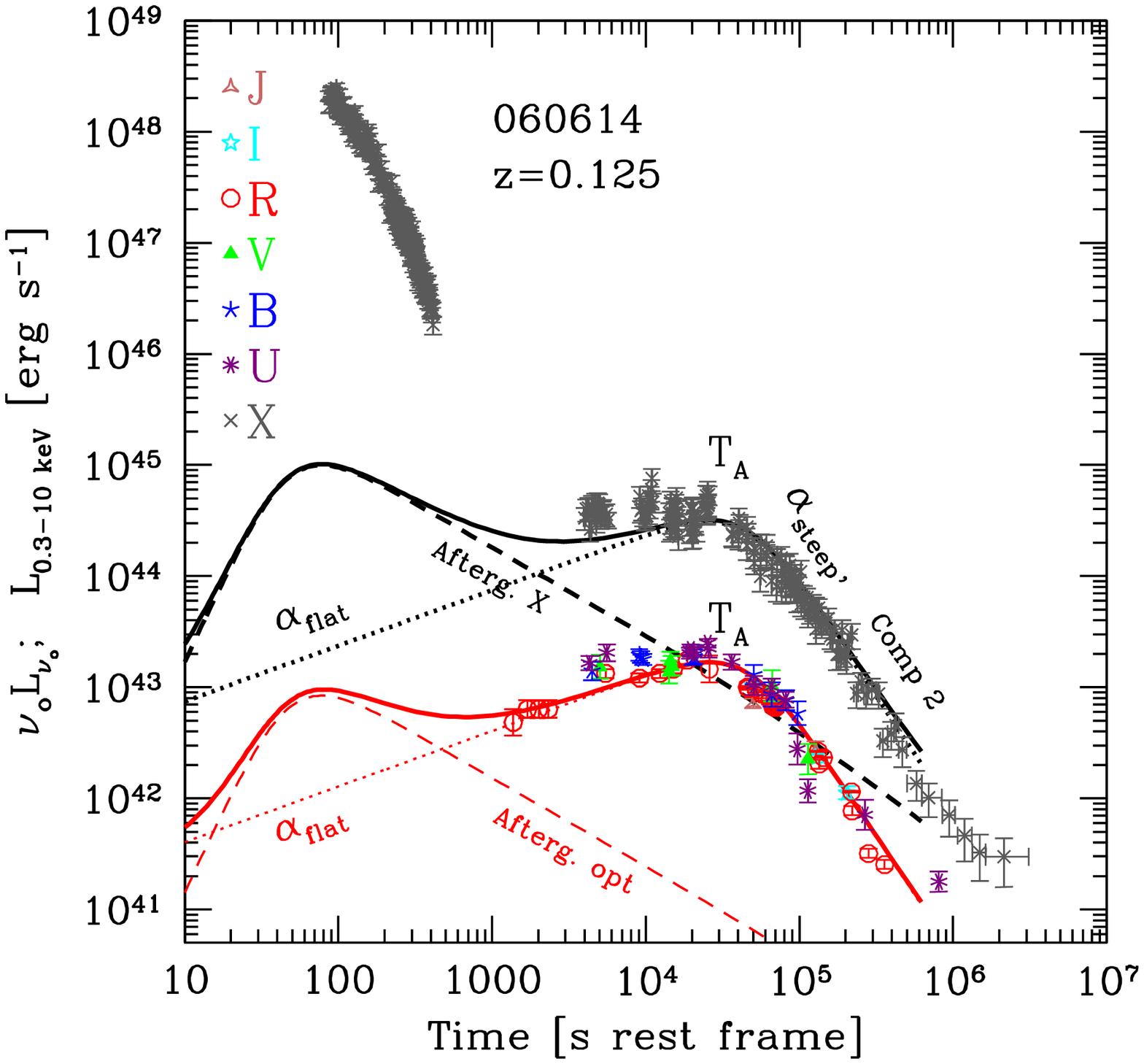}    
  \caption{The X--ray (grey, upper data) and optical 
(coloured, lower data) light--curves of GRB 061126 (left) and GRB 060614 (right).
They are shown in the form of luminosity vs rest frame time.
As can be seen, the optical and the X--ray light--curves of GRB 061126 behave
differently, while they track one another in the case of GRB 060614.
The dotted lines show the emission from Component 2, while the dashed lines 
show the emission of the standard external shock component.
The solid thick line is their sum.
For GRB 061126, the vertical line marks the expected jet--break time,
if the burst follows the ``Ghirlanda'' relation (\cite{ggl04}, \cite{ghirla07}).
In both cases the light--curves at late times are dominated by Component 2, 
while the standard afterglow dominates only at very early times.
At the time of the predicted jet break, GRB 061126 is dominated by Component 2 in 
both bands, and therefore the jet break cannot be seen (see also \cite{marco09}). 
Adapted from \cite{gg09}.
}
\label{f1}
\end{figure}

Specific models often focused only on the X--ray behaviour,
but there is another crucial information that cannot be overlooked:
{\it the X--ray and optical light--curves often do not 
track one another} (e.g. \cite{pana06, pana06b}),
in a way that cannot be accommodated by the standard external 
shock/fireball scenario.
It is this remarkable characteristic that suggests that there must be 
two components contributing to the observed flux.
Fig. \ref{f1} shows two examples of optical and X--ray afterglows,
to illustrate the different behaviour of the optical and X--ray light--curves
in GRB 061126, while they track one another in GRB 060614.

We (\cite{gg09}, see also \cite{marco09}) then used a simple phenomenological 
2--component model (briefly described in the following section) to fit {\it both} the X--ray and 
optical light--curves of {\it Swift} GRBs with redshift and a well sampled optical 
light--curve, allowing to estimate the optical extinction due to dust in the host galaxies. 
As of March 2008, they amount to 33 GRBs.

The logic we follow is to see if with a minimum number of parameters
we can fit the observed light--curves of these bursts. 
If so, we can find some hints for a more physical interpretation.

\section{A two--component model}

We assume that at all times the flux is the sum of two
components: a standard afterglow plus (for the time being) a
completely phenomenological second component.

{\bf Component 1:} this is the synchrotron radiation produced by the
standard forward shock caused by the fireball running into the
circum--burst material.
For this component we follow the prescription of \cite{pana2000},
that requires 6 free parameters, plus the assumption of
an homogeneous or a wind--like profile of the circumburst density.
These are:
i) the isotropic equivalent kinetic energy of the fireball $E_0$;
ii) its initial bulk Lorentz factor $\Gamma_0$; 
iii) the value of the circum--burst medium density $n_0$;
iv)--v) the ``equipartition'' parameters $\epsilon_{\rm e}$ and
$\epsilon_{\rm B}$;
vi) the slope of the relativistic electron energy distribution $p$.

{\bf Component 2:}
it is treated phenomenologically,
since its form/origin is not currently known, though it can be possibly
ascribed to the extension in time of the early prompt emission (as discussed
below).
We parametrise this component with the only criterion of minimising 
the number of free parameters. 

For simplicity we assume that its
spectral shape is constant in time and is described by a 
broken power--law, with spectral indices $\beta_{\rm o}$
and $\beta_{\rm x}$ below and above the break frequency $\nu_{\rm b}$.

The temporal parameters are described by the flat and steep decay
indices, $\alpha_{\rm flat}$ and $\alpha_{\rm steep}$ respectively, and the
time $T_{\rm A}$ at which the two behaviours join.
The time $T_{\rm A}$ is the time at which the shallow phase
(of the steep--flat--steep behaviour) ends.

Then we have 3 spectral and 3 temporal parameters, and we must add one
normalisation. 
Component 2 is thus determined by 7 free parameters, 
that can be rather well constrained by observations 
when this component dominates the emission.
In this case $\alpha_{\rm flat}$, $\alpha_{\rm steep}$ and $T_{\rm A}$ 
can be directly determined, as well as one spectral index
(usually $\beta_{\rm X}$, since the late prompt emission is usually
dominating in the X--ray range).

\begin{figure}
\includegraphics[height=8.5cm]{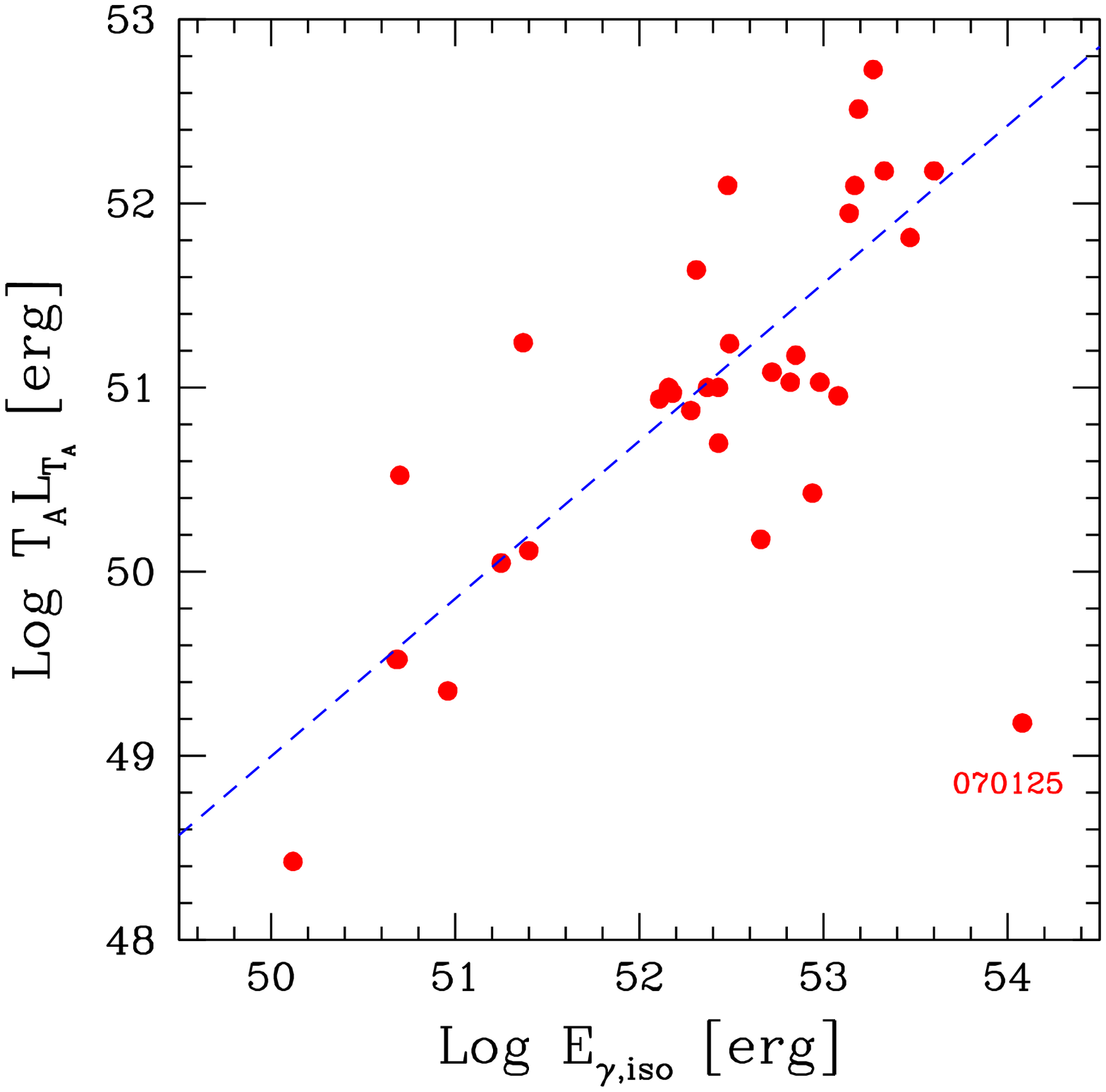}  
\includegraphics[height=8.5cm]{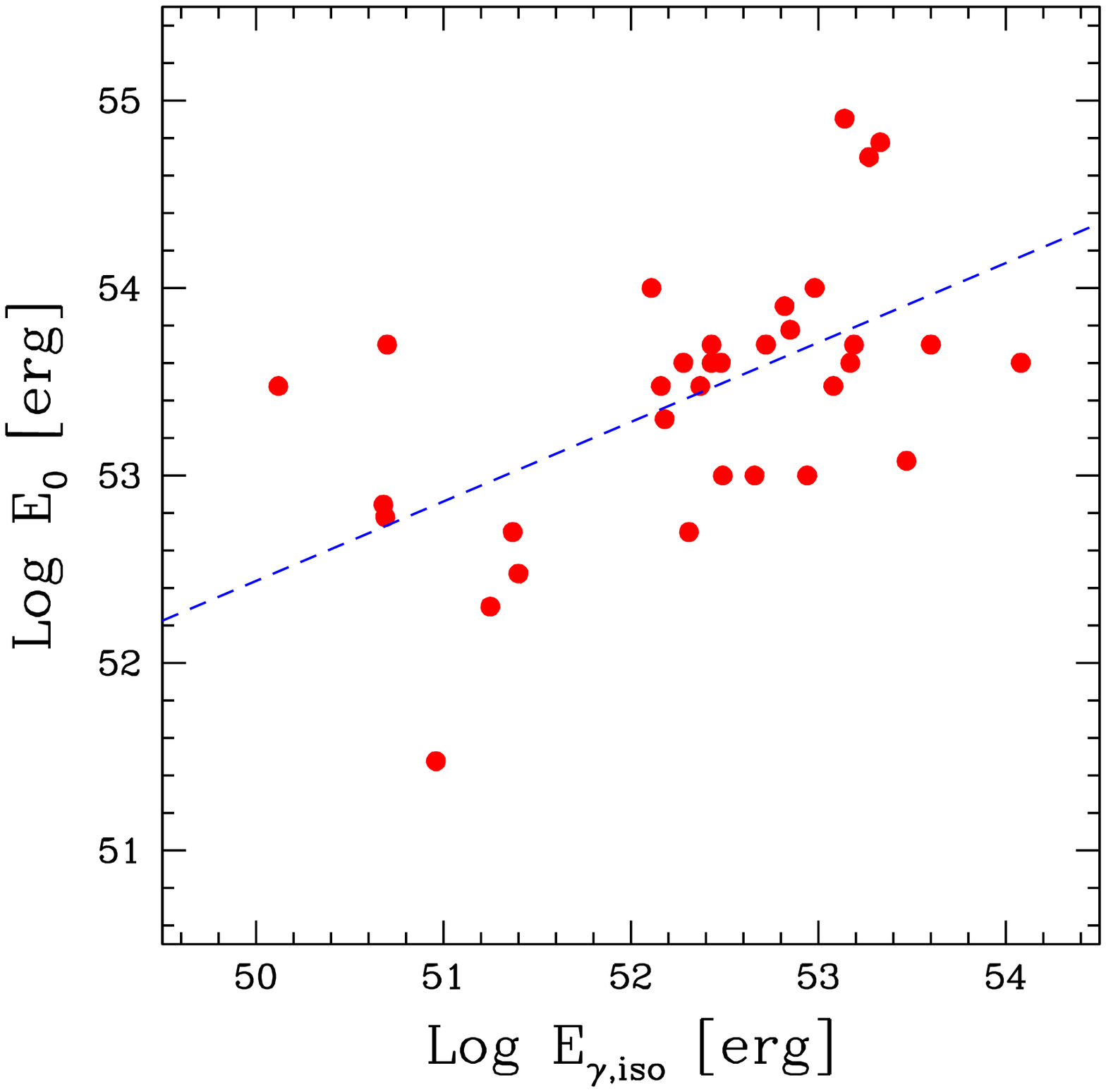} 
\caption{
Left panel: Energy of the late prompt emission, estimated as 
$T_{\rm A} L_{T_{\rm A} } $, as a function of the isotropic energy 
of the prompt emission, $E_{\rm \gamma, iso}$.  
The dashed line corresponds to the least square fit,
$ [T_{\rm A} L_{T_{\rm A}} ] \propto E_{\rm \gamma, iso}^{0.86}$
(chance probability $P= 2\times 10^{-7}$, excluding the outlier GRB 070125).
%
%
Right panel:
the kinetic energy $E_0$ after the prompt emission
as a function of $E_{\rm \gamma,iso}$.  
The dashed line is a least square fit, yielding 
$E_0 \propto E_{\rm \gamma,iso}^{0.42}$
(chance probability $P\sim 10^{-3}$, excluding GRB 070125). 
Adapted from \cite{gg09}.
%
%
}
\label{f2}
\end{figure}

\section{Results}

Fig. \ref{f1} shows two examples of our modelling (in \cite{gg09} one can find all
the 33 GRBs we have analysed).
In these two illustrative examples both GRBs are dominated by Component 2
at late times in both bands, while the standard afterglow emission (i.e. the
synchrotron emission from the external shock) is important only at very early times.
The complex shape of the optical light--curve of GRB 061126 is due to the emerging
of Component 2  at $t\sim$1000 s. 
The break at $t\sim 3000$ s has nothing to do with a jet break, but is the end
of the shallow component (i.e. it is $T_{\rm A}$, the time break of Component 2,
see \cite{nava07} for more discussion).
If this burst (of known $E_{\rm peak}$) had to follow the ``Ghirlanda" relation, then 
it should have a jet break a few days after trigger (rest frame time) as indicated by 
the vertical line.
But at this time the emission is dominated by component 2, {\it thus the jet break
is hidden, and cannot be observed}.
See also \cite{marco09} for more examples and discussion of this point.
For GRB 060614 the entire observed light--curves are dominated
by Component 2: also in this burst the very prominent and {\it achromatic} break at 
$t\sim$40,000 s (rest frame) has again nothing to do with a jet break, being
associated to $T_{\rm A}$.
Note that in our modelling we do not treat the steep phase immediately following
(in most, but not all GRBs) the prompt phase, nor the X--ray or optical flares.
Consider also that the standard afterglow code we use assume isotropic emission,
and so it cannot produce any jet break (i.e. for simplicity 
we do not use the opening angle of the jet as a free parameter). 

\begin{figure}
\includegraphics[height=8cm,width=8cm]{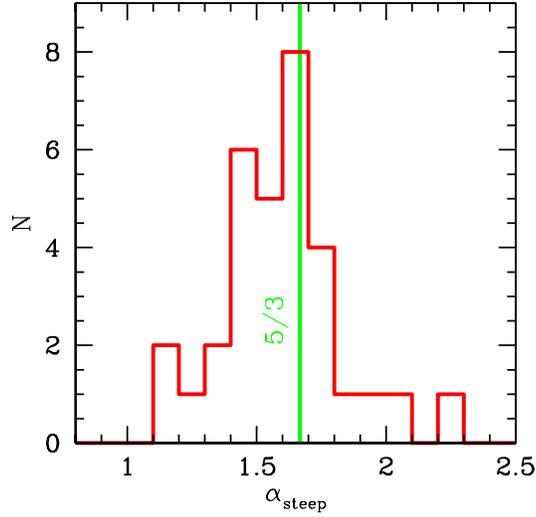}  
\caption{
The distribution of the steep decay index of Component 2.
There is a clustering around the 5/3 value.
This slope coincides with the time profile of the accretion rate
of fallback material in supernovae, that can last for weeks. Adapted from \cite{gg09}.
}
\label{f3}
\end{figure}

\begin{figure}
\includegraphics[height=8.5cm]{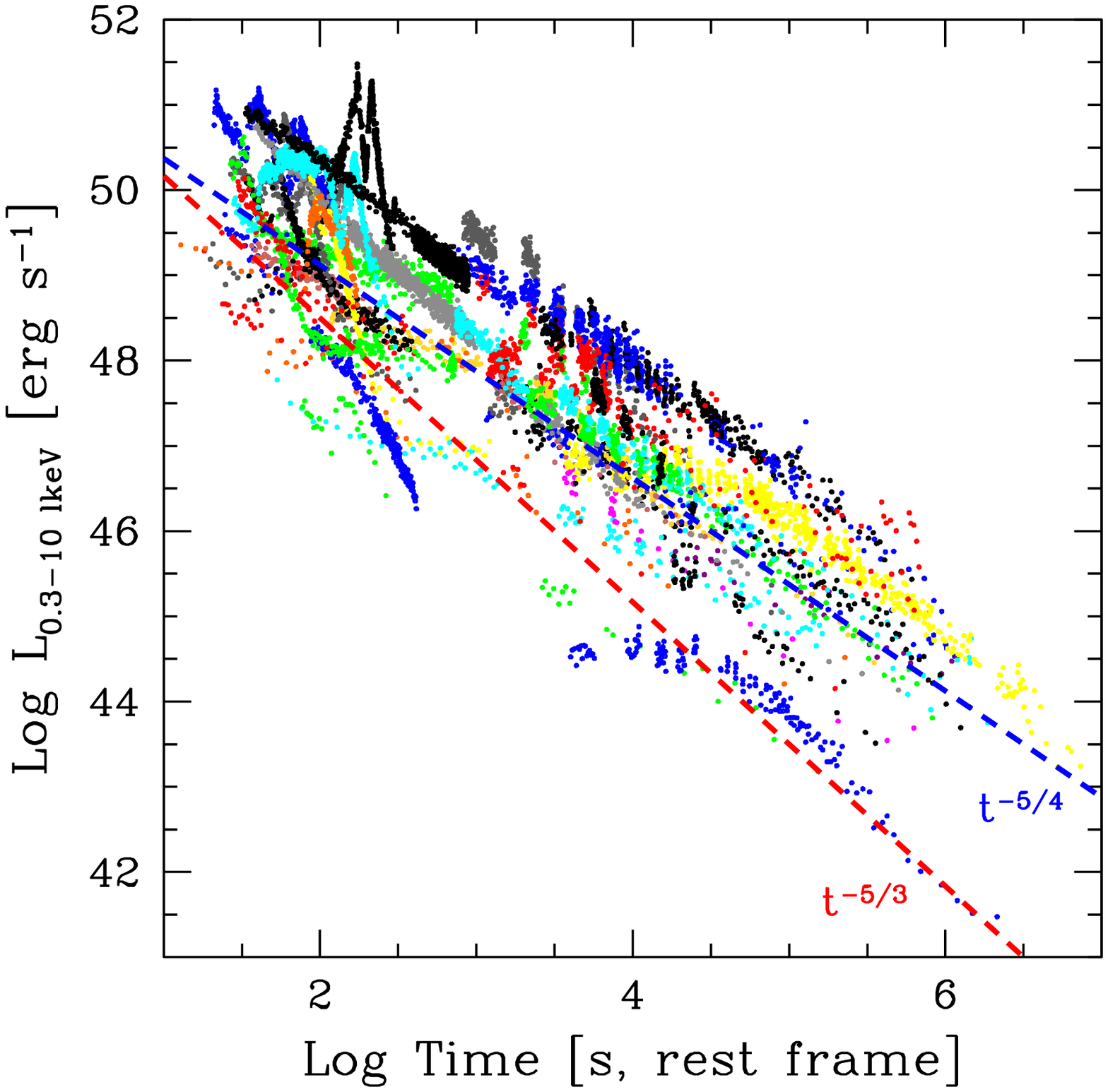}  
\includegraphics[height=8.5cm]{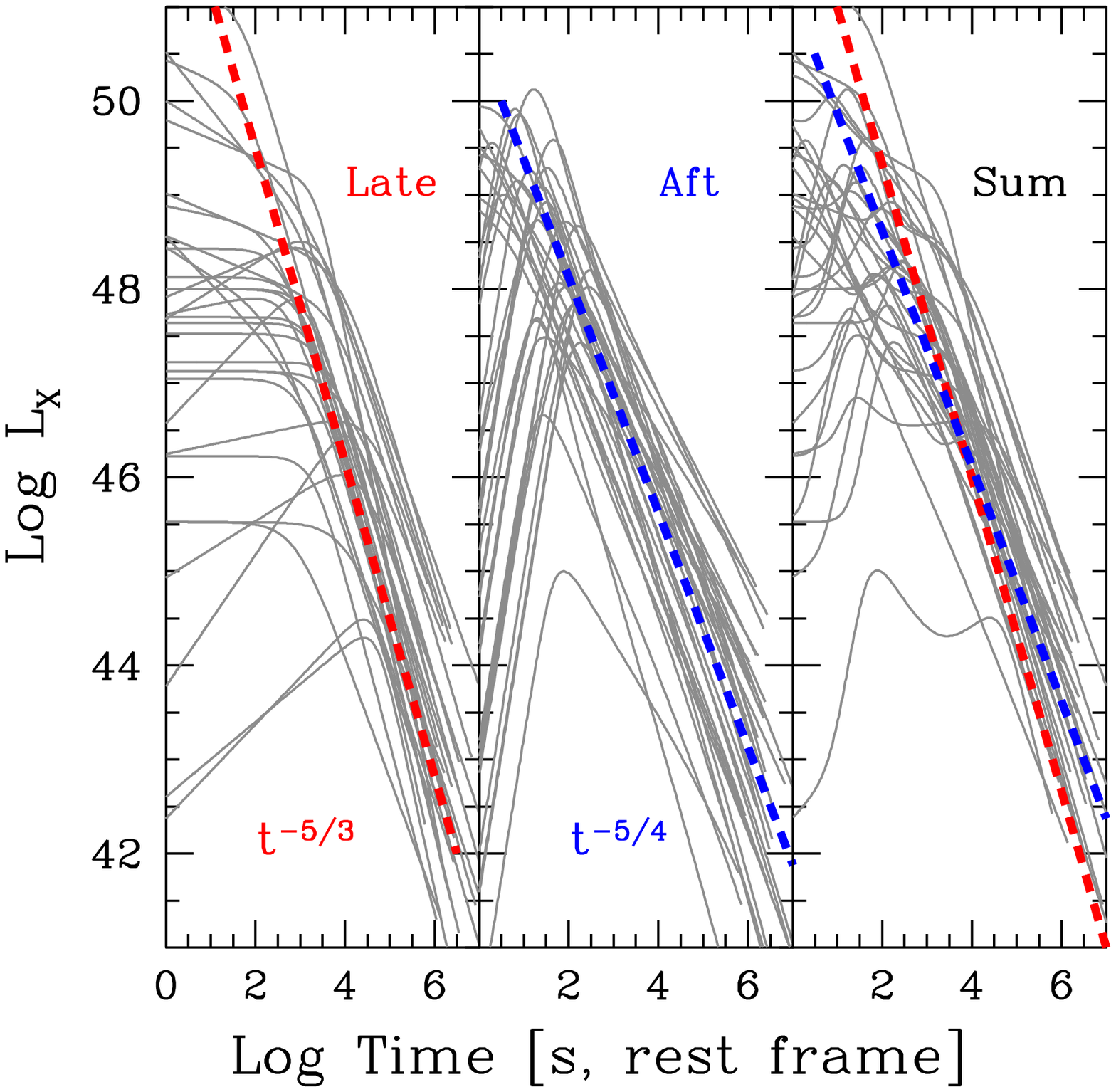}  
\caption{
Left panel:
The light curves of all the 33 GRBs in the X--rays.  
For comparison, the dashed lines
correspond to $t^{-5/4}$ and $t^{-5/3}$, as labelled.
The X--ray luminosity profile is flatter than $t^{-5/3}$
and closer to a $t^{-5/4}$ decay.  
However, this behaviour is due to
the contribution in some GRBs of the afterglow emission at late times,
flattening the overall light curve. 
In fact the right panel shows the modelled X--ray light curves.
On the left we show Component 2 (i.e. the ``late prompt"), in the middle
the afterglow and on the right their sum.
The  dashed lines correspond to $L\propto t^{-5/3}$ or 
$L\propto t^{-5/4}$, as labelled. Adapted from \cite{gg09}.
}
\label{f4}
\end{figure}

Considering now the entire sample of 33 GRBs, we can see if there is
some relation between the (kinetic) energy of the fireball $E_0$
and the energy emitted during the prompt phase (i.e. $E_{\rm \gamma, iso}$).
Similarly, we can see how the energy emitted by Component 2,
approximated by $T_{\rm A}L_{T_{\rm A}}$,
relates with $E_{\rm \gamma, iso}$.
We are aware of the danger of finding signs of a correlation when 
considering two energies (or luminosities), since both quantities are function of redshift,  
but we can compare the two plots. 
Fig. \ref{f2} shows that $T_{\rm A}L_{T_{\rm A}}$ correlates 
with $E_{\rm \gamma, iso}$ better than $E_0$.
This suggests that the early prompt and Component 2 are related. 

The most important results of our analysis is the found 
distribution of $\alpha_{\rm steep}$, the decay index of Component 2
after $T_{\rm A}$.
Fig. \ref{f3} shows a clustering around $\alpha_{\rm steep}=5/3$.
This value is equal to the time profile of the accretion rate of
fallback material in supernovae
(\cite{chevalier89}, \cite{macfadyen01}, \cite{zhang07}), 
and to the average decay of the X--ray flare luminosity, 
as analysed by \cite{lazzati08}.

This is {\it not} the average observed decay slope:
the X--ray light--curves are flatter than 
$L(t)\propto t^{-5/3}$ (see Fig. \ref{f4}), but this is due to the
contribution, especially at early and late times,
of the emission of the standard afterglow.  
The light--curves of Component 2 are indeed steeper, on average, 
than the total emission that reproduces the data. 

The fact that $\langle\alpha_{\rm steep}\rangle\sim 5/3$ strongly
suggests that Component 2 can be interpreted as due 
to the late time accretion  of fallback mass, 
namely material that failed to reach the
escape velocity from the exploding progenitor star, and falls back.
This can continue for weeks, enough to
sustain late prompt emission even at very late times.  
Furthermore, the fact that also X--ray flares follow a similar behaviour
suggests that both X--ray flares and Component 2 have a common origin, 
related to the accretion of the fallback material.

\section{Discussion}

The puzzling features of the early afterglow disclosed by
{\it Swift} have revealed an unforeseen complexity, difficult
to explain in terms of the standard internal/external shock scenario.
Many alternatives have been proposed (see the introduction of \cite{gg09} 
for a brief list of models and \cite{zhang07a} for a review), but only a few 
are able to explain a different behaviour of the optical and X--ray 
light--curves.
Uhm \& Beloborodov \cite{uhm07} and Genet, Daigne \& Mochkovitch 
\cite{genet07} suggested that
the X--ray plateau emission is not due to the
forward, but to the reverse shock running into ejecta of
relatively small (and decreasing) Lorentz factors.
The optical can instead be due to the standard emission of the forward shock.
This scenario requires an appropriate $\Gamma$--distribution of the
ejecta, and also the suppression of the X--ray flux produced by the 
forward shock.

We (\cite{gg07}) instead suggested  
that the plateau phase of the X--ray emission (and sometimes even of 
the optical) is due to a prolonged activity of the central engine (see also 
\cite{lp07}), responsible for a ``late--prompt'' phase: 
after the early ``standard" prompt the central engine continues to 
produce for a long time (i.e. days) shells of progressively lower 
power and bulk Lorentz factor.
The dissipation process during this and the early phases 
occur at similar radii (namely close to the transparency radius).
The reason for the shallow decay phase, and for the break ending it,
is that the $\Gamma$--factors of the late shells are
monotonically decreasing, allowing to see an
increasing portion of the emitting surface, until all of it is visible.
Then the break occurs when $\Gamma=1/\theta_j$, at $T_{\rm A}$.
The shallow phase is then the result of a balance:
the total emitted luminosity is decreasing (with a decay slope $\alpha_{\rm steep}$),
but for some time (before $T_{\rm A}$)
the surface visible to us is increasing (because of the decreasing $\Gamma$).
The combination of these two effects flattens the observed flux decay, 
that appears to be characterised by $\alpha_{\rm flat}$.
After $T_{\rm A}$ the entire emitting surface is visible, and we see the ``real"
decay slope $\alpha_{\rm steep}$.
With respect to other models, this ``late prompt" scenario is
very economic, since the total extra energy needed 
is a relatively small fraction of what used during the real prompt phase.

Finally, we would like to collect the different pieces of evidence 
found in these studies, to try to construct an heuristic simple scenario.
Let us assume that, following the death of a massive star, a rapidly spinning
black hole is formed.
The fast rotation of the equatorial material prevented it to immediately fall
into the black hole.
This material forms a very dense torus.
Viscosity and angular momentum conservation makes this torus to spread, and
accretion takes place.
This accretion phase corresponds to the real prompt emission, possibly mediated
by the strong magnetic field formed in the vicinity
of the black hole, making the Blandford \& Znajek \cite{bz} mechanism at work.
The energy stored in a maximally spinning black hole of 2 solar masses, 
and that can be extracted, amounts to $10^{54}$ erg.
We need only a few per cent of it.
When the bulk of the dense torus has been accreted, there must be a
discontinuity in the accretion rate, to explain the steep 
behaviour of the light--curve following the real prompt emission.
This discontinuity could be associated to the transition from the 
end of the accretion of the material of the dense
torus and the beginning of the accretion of the fallback material.
After this transition phase accretion proceeds at a reduced rate: this phase is
associated to fallback, with its typical time profile.
A reduced accretion most likely corresponds to a reduced magnetic field
in the vicinity of the black hole, and so to a reduced capacity to
extract the spin energy of the black hole. 
Occasionally, however, fragmentation of the accreting material leads to
temporary enhanced accretion, that we can associate to the X--ray flares.


\begin{theacknowledgments}
We acknowledge ASI for I/088/06/0 contract and a 2007 INAF--PRIN grant
for partial funding. 
We are grateful to the Scientific Office of the Italian 
Embassy in Cairo for the financial support. 
\end{theacknowledgments}


\end{document}